\documentclass[showpacs,showkeys,prc,twocolumn]{revtex4}
\usepackage{amssymb}
\usepackage{amsmath}
\usepackage[dvipdfm]{graphicx}
\usepackage{dcolumn}
\usepackage{bm}
\usepackage{amsfonts}

\begin{document}

\title{Entrance-channel dependence of fission transients.}
\author{R. J.~Charity}
\affiliation{Department of Chemistry, Washington University, St. Louis,\\
Missouri 63130}

\begin{abstract}
Fission transients describe the fission rate as it evolves towards the
quasistationary value given by Kramers' formula. The nature of fission
transients is dependent on the assumed initial distribution of the compound
nuclei along the fission coordinate. Although the standard initial
assumption of a near-spherical object leads to a transient suppression of
the fission rate (fission delay), a moderate initial fissionlike deformation
can reduce the magnitude of this suppression. For still larger initial
deformations, transient fission enhancements are possible. Examples of this
behavior are illustrated via a one-dimensional Langevin simulation. It is
argued that the initial conditions are determined by the fusion dynamics and
thus are entrance-channel dependent. Transient fission may be considered
intermediate between statistical fission and quasifission as the rapid time
scale of transient fission may not lead to an equilibrium of the angular and
mass-asymmetry coordinates. The relationship between the mean first passage
time and the transients are discussed. For temperatures much smaller than
the fission barrier, the mean first passage time is independent of the
nature of the fission transients if there is no strong competition from
evaporation. Thus, fission transients are most important when the
evaporation time scale is smaller than, or of the order of, the transient
time.
\end{abstract}

\maketitle

\section{INTRODUCTION}

Nuclear fission at high excitation energies can be modeled as a
diffusion-driven escape over a potential barrier. The quasi-stationary or
equilibrium fission rate for a one-dimensional problem was determined by
Kramers\cite{Kramers40} in 1940. Forty years later, Grang\'{e} and Weidenm%
\"{u}ller\cite{Grange80} considered the time dependence of the fission rate
as it approaches this equilibrium limit. Of course, the nature of this
transient rate depends on the assumed initial distribution along the fission
coordinate. Grang\'{e} and Weidenm\"{u}ller and subsequently almost all
other treatments of this problem\cite%
{Grange83,Hassani84,Weidenmuller84,Grange84,Bhatt86,Grange86,Zhongdao86,Delagrange86,Frobrich93,Gonchar95,Abe96,Chaudhuri01}
assumed that all systems start out near-spherical or with\ deformations
corresponding to the local minimum in the potential energy surface. As a
consequence, it takes time for the population at the higher deformations
near the saddle point to build up and therefore the fission rate is
initially suppressed compared to the equilibrium value. This has led to the
concept of a fission delay.

If the light-particle evaporation time scale is shorter than the transient
or delay time $\tau _{trans}$, then one would expect an increased
multiplicity of evaporated particles before the system is committed to
fission. It has been suggested that this contributes to the large number of
neutrons and other particles which are emitted prior to scission\cite%
{Hilscher92}. However, these prescission particles also have contributions
associated with evaporative emissions as the fissioning system descends from
the saddle to the scission point. A full understanding of the relative
contributions from presaddle and saddle-to-scission emissions has not yet
been obtained.

It is the purpose of this work to illustrate the sensitivity of the fission
transients to the assumed initial distribution using a schematic
one-dimensional model. Most calculations will be performed for the limit of
high friction which allows a number of simplifications in the calculations.
Also a number of studies have suggested this friction limit is appropriate
for fission\cite{Hinde86,Hinde89}.

\section{LANGEVIN SIMULATIONS}

Consider the motion along the fission coordinate $x$ subjected to a
potential energy $V(x)$, of the form indicated in Fig.~1, with a local
minimum or ground-state\ at $x$=$x_{\min }$ and a local maximum or
saddle-point at $x$=$x_{\max }$. The fission barrier is therefore $B$ $%
=V(x_{\max })$ $-V(x_{\min })$. The one-dimensional Langevin equation for
this system is 
\begin{equation}
M\ddot{x}=-dV/dx-\gamma \dot{x}+\xi \left( t\right)
\end{equation}%
where $M$ is the inertia, $\gamma $ is the friction, and $t$ is time. The
fluctuating Langevin force $\xi \left( t\right) $ is dependent on the
temperature $T$ of the system and its time dependence has the following
properties 
\begin{equation}
\left\langle \xi \left( t\right) \right\rangle =0,\;\left\langle \xi \left(
t\right) \,\xi \left( t^{\prime }\right) \right\rangle =2\gamma T\,\delta
\left( t-t^{\prime }\right) .
\end{equation}%
In the limit of high friction, the inertial term drops out and the reduced
Langevin equation is 
\begin{equation}
\dot{x}=-\frac{V^{\prime }(x)}{\gamma }+\frac{\xi \left( t\right) }{\gamma }.
\end{equation}

Consider the motion of the system inside the potential well without the
fluctuating force, i.e. , $\xi \left( t\right) $=0. In this case for all
initial deformations not to far from $x_{\min }$, the solution of this
equation is just an exponential approach of the deformation towards $x_{\min
}$ with a time constant of $\tau =\gamma /\left( \frac{d^{2}V}{dx^{2}}%
\right) _{x=x_{\min }}$. This time constant set the time scale for motion
inside the well.

\begin{figure}[tbp]
\includegraphics*[scale=.4]{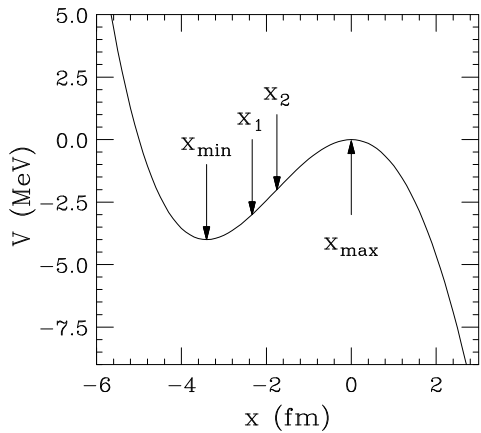}
\caption{Variation of the potential energy $V$ with the fission coordinate $x
$ used in the Langevin simulations. The locations of the three initial
starting configurations are indicated as $x_{\min }$, $x_{1}$, and $x_{2}$. }
\end{figure}
The quasi-stationary or equilibrium decay rate in the limit of high friction 
\cite{Kramers40} is given by Kramers formula; 
\begin{equation}
k_{f}^{\infty }=\frac{\sqrt{\left( \frac{d^{2}V}{dx^{2}}\right) _{x=x_{\min
}}\;\left\vert \left( \frac{d^{2}V}{dx^{2}}\right) _{x=x_{\max }}\right\vert 
}}{2\pi \gamma }\exp \left( \frac{-B}{T}\right) \left( 1-\lambda T\right) .
\label{Eq:Kramers}
\end{equation}%
This formula is strictly valid for $T\ll B$. The last term contains the
first-order correction in $T$ from Ref.~\cite{Edholm79} where 
\begin{eqnarray}
\lambda  &=&\frac{\left( \frac{d^{4}V}{dx^{4}}\right) _{x=x_{\max }}}{8\,%
\left[ \left( \frac{d^{2}V}{dx^{2}}\right) _{x=x_{\max }}\right] ^{2}}-\frac{%
\left( \frac{d^{4}V}{dx^{4}}\right) _{x=x_{\min }}}{8\,\left[ \left( \frac{%
d^{2}V}{dx^{2}}\right) _{x=x_{\min }}\right] ^{2}}  \notag \\
&&+\frac{5\,\left[ \left( \frac{d^{3}V}{dx^{3}}\right) _{x=x_{\max }}\right]
^{2}}{24\,\left\vert \left( \frac{d^{2}V}{dx^{2}}\right) _{x=x_{\max
}}\right\vert ^{3}}+\frac{5\,\left[ \left( \frac{d^{3}V}{dx^{3}}\right)
_{x=x_{\min }}\right] ^{2}}{24\,\left[ \left( \frac{d^{2}V}{dx^{2}}\right)
_{x=x_{\min }}\right] ^{3}}.
\end{eqnarray}

In order to determine the transient fission decay rate, the Langevin
equation is solved in a Monte Carlo fashion starting at some initial value $%
x_{0}$ of the fission coordinate. Each Langevin trajectory can be followed
until it crosses the saddle point at $x_{\max }$. However, the system is not
committed to fission at this point, the fluctuating force can drive the
system back across the saddle point towards $x_{\min }$. The system only
becomes completely committed to fission once the scission point is reached.
However in practice once the system has crossed the saddle point and
descended to a point $x$ where $V(x_{\max })-V(x)>2T$, the possibility of
returning to the saddle point is remote. So the Langevin trajectories are
followed to an effective scission point $x_{e.s.}$ where $V(x_{\max
})-V(x_{e.s.})=2T$. However in order to define a fission time independent of
this or the real scission point, the time of the last passage through the
saddle point is chosen. Note, if the real scission point is closer to $%
x_{\max }$ than our effective saddle point $x_{e.s.}$, then the equilibrium
rate also needs to be modified. From Fr\"{o}brich, Gontchar, and Mavlitov 
\cite{Frobrich93}, when the saddle and scission points are close, the
quasistationary decay rate is 
\begin{equation}
k_{f}^{\infty }\approx k_{f}^{\infty }(\mathrm{Kramers})\,\frac{2}{\left\{ 1+%
\mathrm{erf}\left[ \sqrt{\frac{V(x_{\max })-V(x_{s})}{T}}\right] \right\} }
\end{equation}%
where $k_{f}^{\infty }(\mathrm{Kramers})$ is Kramers' formula (Eq.~\ref%
{Eq:Kramers}). If the saddle and scission points are identical ($x_{\max }$ $%
=x_{s}$), then fission rate is a factor of two larger than Kramers' value.
It rapidly approaches $k_{f}^{\infty }(\mathrm{Kramers})$ when the scission
point is further away from the saddle point. Note for our effective scission
point ($x_{s}=x_{e.s.}$), this formula gives a fission rate which is only
2.3\% larger than $k_{f}^{\infty }(\mathrm{Kramers})$ and this small
enhancement can be ignored.

An alternative to the Langevin simulations is to solve the Fokker-Planck or
Smoluchowski equations for the probability distribution function $P(x,t)$.
Grang\'{e} and Weidenm\"{u}ller\cite{Grange80} define the transient fission
rate in their Fokker-Planck calculations as 
\begin{equation}
\frac{dN}{dt}=-k_{f}(t)\,N,\;N=\int_{-\infty }^{x_{\max }}P(x,t)\,dx.
\label{Eq:rate_w}
\end{equation}%
This definition is used in Ref.~\cite{Grange83} and more recently in Ref.~%
\cite{Jurado03}. With this definition, those systems that have crossed the
saddle point at $x_{\max }$ are no longer counted as part of the
\textquotedblleft ground-state\textquotedblright\ population $N$ and the
fission rate is equated to the rate of decay of this population. However as
already mentioned, those systems that have just crossed the saddle point are
not necessarily committed to fission and have a possibility of rejoining
this population. Thus, this definition is not equivalent to that adopted is
this work and leads to smaller transient times.

Calculations were performed with the potential energy which was first used
in Ref.~ \cite{Grange83}, and subsequently in Refs.~\cite{Grange84,Bhatt86}.
The potential, shown in Fig.~1, was deemed appropriate for a compound
nucleus with $A=248$ and has a fission barrier of $B$=4 MeV. A friction
parameter of $\gamma =$3.4~zs MeV/fm$^{2}$ was assumed. The transient
fission rate relative to the equilibrium value [Eq.~(\ref{Eq:Kramers})] is
plotted against $t/\tau $ in Fig.~2 for $T$=1~MeV and with the standard
initial condition of $x_{0}=x_{\min }$. The purpose of plotting the results
in this particular way is that they are independent of the friction
parameter $\gamma $. Weidenm\"{u}ller and Jing-Shang\cite{Weidenmuller84}
first pointed out the simple scaling of the fission rate with friction in
the high-friction limit, i.e., $k_{f}(t,\gamma _{1})$ $=(\gamma _{2}/\gamma
_{1})k_{f}(\gamma _{1}t/\gamma _{2},\gamma _{2})$. Of course the friction
must be large enough that the limit is approached, but otherwise the plotted
results are independent of $\gamma $. The results obtained with the
definition of the fission time proposed in this work and from the definition
of Grang\'{e} and Weidenm\"{u}ller [Eq.~(\ref{Eq:rate_w})] are indicated by
the filled and open data points, respectively. With both definitions, the
fission rate is initially suppressed as expected for $x_{0}=x_{\min }$ and,
as also expected, the transient time is a little shorter with the definition
of Weidenm\"{u}ller and Jing-Shang. The dashed curve indicates a form of the
transient rate which has found some use in statistical-model calculations 
\cite{Dioszegi00,Shaw00,Dioszegi01}, 
\begin{equation}
k_{f}(t)=k_{f}^{\infty }\left[ 1-\exp \left( -\frac{2.3t}{\tau _{d}}\right) %
\right] ,\;\tau _{d}=\frac{\tau }{2}\ln \left( \frac{10B}{T}\right) .
\label{eq:tau}
\end{equation}%
One can clearly see it is not a very good approximation. 
\begin{figure}[tbp]
\includegraphics*[scale=.4]{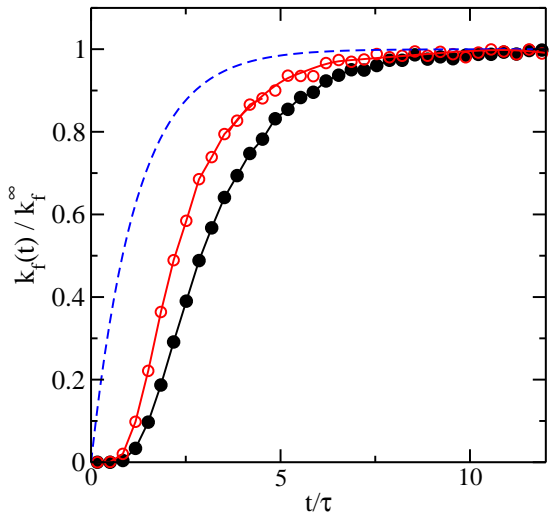}
\caption{(Color online) The transient fission decay rates, normalized
relative to the equilibrium rate, are plotted versus the normalized time $t/%
\protect\tau $. The results, indicated by the solid data points, were
obtained from the definition of fission time adopted in this work. The open
symbols represent the transient rate as defined by Eq.~(7), while the dashed
curve is from Eq.~\protect\ref{eq:tau}.}
\label{fig:fig2}
\end{figure}

\begin{figure}[tbp]
\includegraphics*[scale=.4]{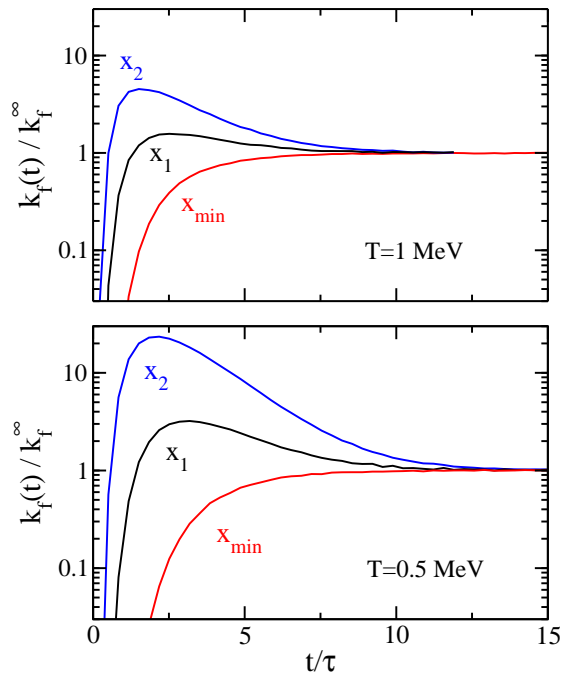}
\caption{(Color online) Normalized transient decay rates as in Fig.~2 for
temperatures of $T$=0.5 and 1~MeV. In each panel, results are shown for the
three indicated initial values of the fission coordinate.}
\label{fig:fig3}
\end{figure}

Let us now concentrate on the dependence of the transients on the initial
deformation. Unlike Ref.~\cite{Weidenmuller84}, in which a dependence on the
initial starting value was also considered, let us consider only
well-bounded initial deformations $x_{0}$ which lie well inside the
potential well, i.e., $B\geq V\left( x_{\max }\right) -V\left( x_{0}\right)
>T$. In such cases, the fission barrier will still be a substantial obstacle
to decay. Figure~\ref{fig:fig3} illustrates the strong dependence of the
transients on the initial condition for two temperatures; $T$= 0.5 and
1~MeV. Three initial deformations $x_{0}=x_{\min }$, $x_{1}$, $x_{2}$ were
considered and their location on the potential energy surface is indicated
in Fig.~1. Note that $V(x_{1})-V(x_{\min })$ = $B/4$ and $V(x_{2})-V(x_{\min
})$ = $B/2$. From Fig.~\ref{fig:fig3}, one can see that for any small
initial fissionlike deformation, such as $x_{1}$, the magnitude of the
fission delay is greatly reduced. For still larger initial deformations,
such as $x_{2}$, one even obtains a substantial transient fission
enhancement. In this case, those systems that fissioned in the transient
period typically did not pass through the ground-state configuration $%
x_{\min }$. The systems which fission first are those for which the
fluctuations drive that system almost continually towards the saddle point.
Such behavior for large initial deformations was initially suggested in Ref.~%
\cite{Hinde86}.

These simulations indicate that the very existence of a fission delay
depends on the appropriate initial deformation. However, all simulations
approach the equilibrium rate over a similar time scale which is associated
with attaining the equilibrium distribution in $x$. Note for smaller
temperatures, the transient deviations from Kramers' rate are larger and the
extent of the transient period is longer. 

\begin{figure}[tbp]
\includegraphics*[scale=.4]{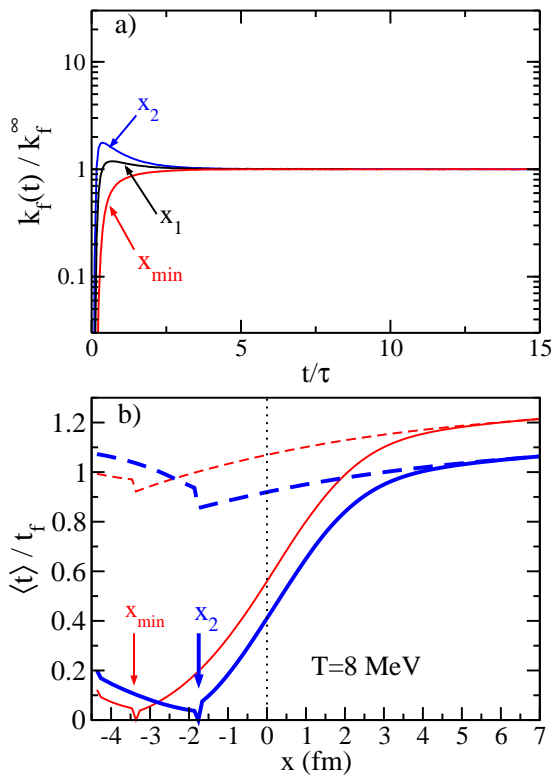}
\caption{(Color online) a) Normalized transient decay rates as in Fig.~2,
but now for a temperature which is larger than the fission barrier, i.e., $T$%
=8~MeV. Results are shown for the three indicated initial values of the
fission coordinate. b) Mean first (solid curves) and last (dashed curves)
passage times obtained in the same simulation. Thick and thin curves
correspond to the two indicated initial deformations.}
\label{fig:t8}
\end{figure}

\begin{figure}[tbp]
\includegraphics*[scale=.4]{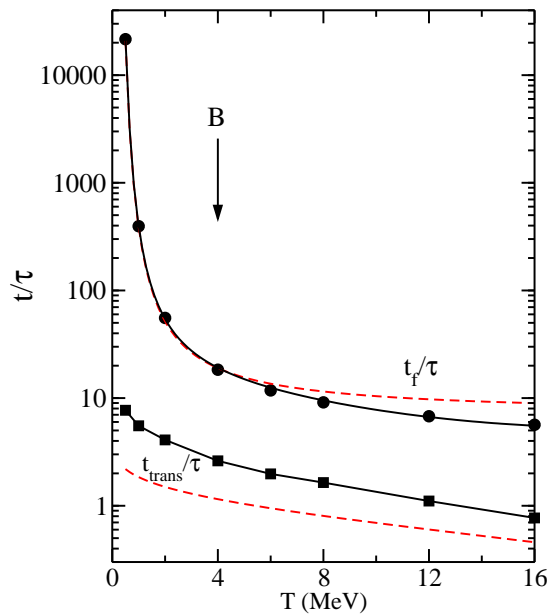}
\caption{(Color online) The mean statistical $t_{f}$ and the transient $%
t_{trans}$ time plotted as functions of the temperature $T$. The data points
were obtained from the Langevin simulations. The solid curves are smooth
fits to these points. The dashed curve for $t_{f}$ is from the Kramers'
formula (without the first order correction). The dashed curve for $%
t_{trans} $ is $\protect\tau _{d}$ from Eq.~\protect\ref{eq:tau}.}
\label{fig:times}
\end{figure}

Although Kramers decay rate is valid only for $T\ll B$, the Langevin
simulations produce qualitatively similar behavior for $T\gtrsim B$. For
example, Fig.~\ref{fig:t8} shows the normalized fission rate for $T=2B$. In
this figure $k_{f}^{\infty }$ is not determined from Eq.~\ref{Eq:Kramers},
but is just the asymptotic value of $k_{f}(t)$ in the simulations. The
transient time scale is much shorter than those in Fig.~\ref{fig:fig3} and
during this period, the deviations of $k_{f}(t)$ from $k_{f}^{\infty }$ are
much smaller. This does not imply transient effects are less important at
higher temperatures. One should remember that the absolute values of $%
k_{f}(t)$ are much larger for the higher temperatures and so the probability
of fissioning during the transient period is much greater for these hotter
systems (ignoring competition for evaporation processes). Let us define the
transient time $t_{trans}$ as the time for $k_{f}(t)$ to attain 90\% of $%
k_{f}^{\infty }$ for $x_{0}=x_{\min }$. The temperature dependence of the
transient time $t_{trans}$ and the mean statistical time $%
t_{f}=1/k_{f}^{\infty }$ are plotted in Fig.~\ref{fig:times} as the data
points. It is important to note that for all temperatures, $t_{trans}<t_{f}$
. Now for $T\ll B$, Fig.~\ref{fig:times} \ shows that $t_{trans}\ll t_{f}$
and thus the number of events that fission during the transient period
approaches zero. Therefore most fission events in these simulations occur
during the period when $k_{f}(t)\sim k_{f}^{\infty }$. 

The dashed curves in Fig.~\ref{fig:times} give the Kramers' formula (without
the first order correction) and $\tau _{d}$ from Eq.~\ref{eq:tau}. Although
Kramers formula is not valid for $T\gtrsim B$, it still predicts the
statistical rate to within an order of magnitude. The quantity $\tau _{d}$
may be used for a similar estimate of the transient time.

\section{MEAN\ FIRST\ AND\ LAST\ PASSAGE\ TIMES}

\begin{figure}[tbp]
\includegraphics*[scale=.4]{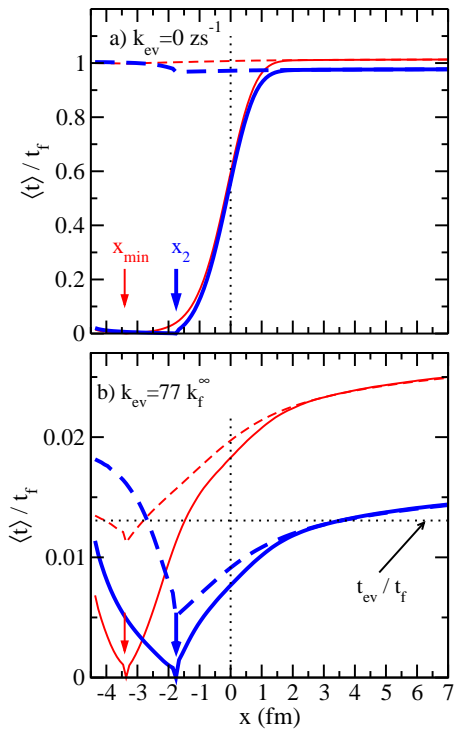}
\caption{(Color online) Mean normalized times plotted as a function of
deformation for $T$=1~MeV. The solid curves show the mean first passage time 
$\protect\tau _{\mathrm{MFPT}\text{ }}$while the dashed curves are for the
mean last passage time $\protect\tau _{\mathrm{MLPT}}$. Results are shown
for two initial deformations indicated by the arrows; $x_{0}=x_{\min }$
(thin curves) and $x_{0}=x_{2}$ (thick curves). \ a) No evaporation
competition. b) Evaporation competition with an evaporation rate of $%
k_{ev}=77\times k_{f}^{\infty }$.}
\label{fig:mftp}
\end{figure}

Hofmann and Ivanyuk\cite{Hofmann03} have recently championed the concept of
the \textit{mean first passage time} $\tau _{\mathrm{MFPT}}$ in fission and
have suggested that fission transients are not important. Let us examine
this closely. The \textit{first passage time} (FPT) in the Langevin
simulations is the first time the deformation passes through a particular
value of $x$. Following Ref.~\cite{Bao04}, the \textit{mean last passage
time }$\tau _{\mathrm{MLPT}}$\textit{\ }can also be considered. The \textit{%
last passage time }(LPT)\textit{\ }is the last time the deformation passes
through a particular value. Note, the fission rate evaluated in this work is
based on the LPT at the saddle point as in Ref.~\cite{Bao04}. Both the $\tau
_{\mathrm{MFPT}}$ (solid curves) and $\tau _{\mathrm{MLPT}}$(dashed curves)
determined for $T$=1~MeV are displayed in Fig.~\ref{fig:mftp}a as a function
of $x$. The two times are normalized relative to the mean statistical time $%
t_{f}$. Results are also shown for two initial deformations, $x_{\min }$
(thin curves) and $x_{2}$(thick curves). As pointed out by Hofmann and
Ivanyuk\cite{Hofmann03}, the value of $\tau _{\mathrm{MFPT}}$\ is not very
sensitive to the initial deformation. The same is also true for the $\tau _{%
\mathrm{MLPT}}$. For deformations well beyond the saddle point ($x$=0), $%
\tau _{\mathrm{MFPT}}\sim t_{f}$ and for all deformations $\tau _{\mathrm{%
MLPT}}\sim t_{f}$. These results are true in general as long as $T/B$ is
small. 

Let us define the thermal length scale as $l_{th}=\sqrt{2T/\left( \frac{%
d^{2}V}{dx^{2}}\right) _{x=x_{\max }}}$. Now in the limit of high friction
and $T/B\rightarrow 0$, H\"{a}nggi \textit{et al.}\cite{Hanggi90} showed
that $\tau _{\mathrm{MFPT}}\rightarrow t_{f}$ for all deformations more than
the distance $l_{th}$ beyond the barrier. This result is independent of
initial deformation as long as this deformation is located before the
barrier by distance greater than $l_{th}$. In the simulations associated
with Fig.~\ref{fig:mftp}, the thermal length is small, i.e., $l_{th}=$1~fm.
The reason for this insensitivity to the initial deformation should be
obvious, the number of events which fission within the transient period is
vanishingly small when $T\ll B$. However for larger temperatures, this is no
longer the case. For example, Fig.~\ref{fig:t8}b shows the mean times for $%
T=2B$. Here there is a more significant dependence on the initial
deformation.

Given the insensitivity of $\tau _{\mathrm{MFPT}}$ and $\tau _{\mathrm{MLPT}%
} $ to the transients effects for $T\ll B$, one may question their relevance
for fission. If all events are allowed to fission as in the previous
simulations, then clearly the transient decay rate has little relevance.
However, fission is not the only decay mode of a compound nucleus.
Competition from light-particle evaporative decays can drastically alter the
probability for fission and make the transients important. Consider
competing fission and evaporation branches. Let $k_{f}\left( t\right) $ now
represent the partial decay rate for fission, i.e., the rate in the absence
of evaporation. Let the partial evaporation rate be $k_{ev}$ and its mean
time, in the absence of fission, is thus $t_{ev}=1/k_{ev}$. If there are no
transients and the partial fission rate is constant, i.e. $%
k_{f}(t)=k_{f}^{\infty }$, then the fission probability is just $%
k_{f}^{\infty }/\left( k_{f}^{\infty }+k_{ev}\right) $ and the final fission
rate is given by the total decay rate $k_{f}^{\infty }+k_{ev}$. If $%
k_{ev}>k_{f}^{\infty }$, then not only is the fission probability small, but
those events that do fission occur on the shorter time scale of $t_{ev}$. If 
$t_{ev}$ is less than, or of the order of, $t_{trans}$, then the transients
will be very important. To illustrate this, the effect of a competing
evaporative decay mode was included in a simplistic way into the Langevin
simulations. For each event, an evaporation time was chosen in a Monte-Carlo
fashion from an exponential distribution with partial decay rate $k_{ev}$.
Each Langevin simulation was then stopped when it reached this time. The
mean times $\tau _{\mathrm{MFPT}}$ and $\tau _{\mathrm{MLPT}}$ were then
determined only from events that reached a given deformation. These are
displayed in Fig.~\ref{fig:mftp}b for the same $T$ and initial deformations
as in Fig.~\ref{fig:mftp}a, but with a mean evaporation time approximately
equal to the transient time, $t_{ev}$=0.93$\times t_{trans}$ ($k_{ev}$=77$%
\times k_{f}^{\infty }$). Clearly from Fig.~\ref{fig:mftp}b, the inclusion
of evaporation competition has now produced a large dependence of the mean
times on the initial deformation. All mean times are much smaller than those
in Fig~\ref{fig:mftp}a and are of the order of the mean evaporation time $%
t_{ev}$. This latter time is indicated in the figure by the horizontal
dashed line. For deformations well beyond the barrier, the times $\tau _{%
\mathrm{MFPT}}$ and $\tau _{\mathrm{MLPT}}$ are identical indicating that
the fluctuating forces are unimportant as the systems smoothly drop down the
barrier. In fact, the relative dependencies of the times in this deformation
region are identical for all curves in Figs.~\ref{fig:mftp}a--\ref{fig:mftp}%
b. This behavior is dictated by the very sharp drop off in the potential
beyond the barrier and is not general for all possible potentials.

The previous simulation where $T\ll B$, $t_{ev}\sim t_{trans}$ and thus $%
k_{f}^{\infty }\ll k_{ev\text{ }}$is not well represented by studies of
prescission particle emission as the small fission probability makes
coincidence measurements with evaporated light particles difficult. For most
measurements with heavy-ion reactions, the fission yield is associated with
compound-nucleus spins where $k_{f}^{\infty }\gtrsim k_{ev}$ and the
sensitivity to the fission transients is not large unless $T\gtrsim B$. \
For example consider the prescission multiplicity measurements of Hinde 
\textit{et al.} for a number of compound systems with $A$=168 to $A$=251\cite%
{Hinde86}. The authors consider a number of simulations including fission
delays and saddle-to-scission evaporations to reproduce the fission and
residue cross sections and the prescission neutron multiplicities. Take for
example the $^{19}$F+$^{181}$Ta reaction producing a compound nucleus with
80~MeV of excitation energy. Most of the fission yield for this reaction
occurs for compound-nucleus spins of $J$=40 to $J$=63~$\hbar $ in their
simulations. In this range, $k_{f}^{\infty }/k_{ev}$ (first chance) varies
from 0.05 to 1.1. If the compound nucleus survives first chance fission
competition, then the surviving daughter nucleus (and its daughters) may
also undergo fission. The total fission yield depends on the contribution
from all chances, i.e., the sum of the fission yield from all daughters
produced in the decay cascade. The total fission probability from all
chances varied from $\sim $56\% to almost 100\% over this spin range, but it
is dependent of the form of the transients assumed. However, only for the
lowest part of this spin range is $k_{f}^{\infty }\ll k_{ev}$ (first chance)
and thus only here may there be a large sensitivity to the transients. In
Ref.~\cite{Hinde86} it was concluded that if all prescission neutrons are
entirely from presaddle emissions, then this would require a transient
fission suppression with $t_{trans}$=70~zs. For the $^{19}$F+$^{181}$Ta
system discussed above, the first-chance statistical-fission time is $t_{f}$%
=70~zs at $J$=45~\ and drops to $t_{f}$=10~zs at $J$=63~$\hbar $. Clearly
such a transient time does not make sense as $t_{f}<t_{trans}$ for these
spin values and this suggests that transients cannot explain all the
observed prescission neutrons in this reaction. Thus, this should serve as a
warning that one cannot assume arbitrary large fission delays when fitting
such data.

\section{WEAK\ FRICTION}

\begin{figure}[tbp]
\includegraphics*[scale=.4]{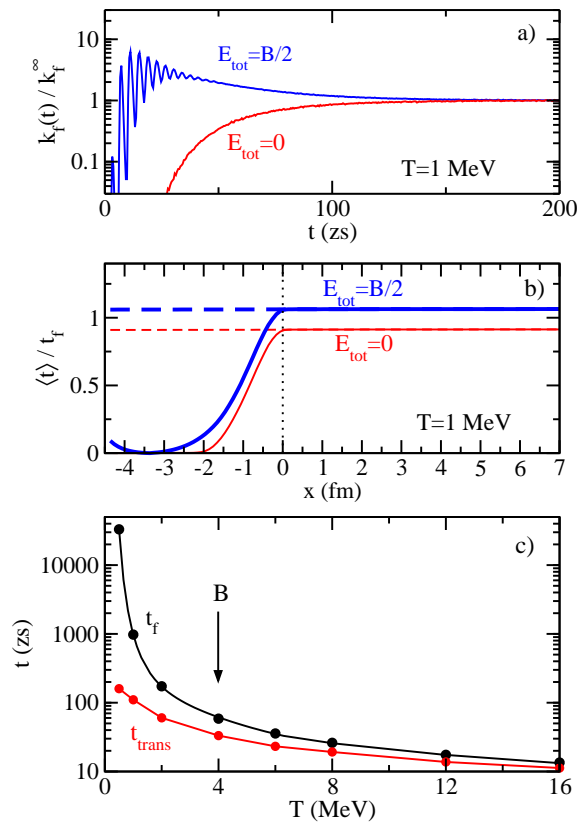}
\caption{(Color online) Results obtained with weak friction. a) Normalized
transient decay rates for the two indicated initial total energies with $T$%
=1~MeV. b) Mean first and last passage times obtained from the same two
simulations as in a). c) Dependence of the mean statistical and transient
times on temperature. The arrow indicates the temperature where $T$=$B$.}
\label{fig:underdamped}
\end{figure}

Although we have considered the high friction limit, most of the previous
conclusions are also valid for moderate to low friction. In general in this
regime, the initial conditions for the simulations must be specified by the
initial deformation and its time derivative. However in the limit of weak
friction, the system will oscillate within the potential well and the total
energy $E_{tot}$ (potential plus kinetic) of the system will evolve more
slowly. Thus in this limit, the initial condition can be specified by the
initial total energy. Results are shown for underdamped simulations in Fig.~%
\ref{fig:underdamped}a. Here the inertia was set to 62 nucleon masses and $%
\gamma =$0.015 ~zs MeV/fm$^{2}$. The reduced friction is thus $\gamma /M$%
=0.023~zs$^{-1}$ and is small compared to the oscillation frequency in the
ground state minimum $\omega =\sqrt{\left( \frac{d^{2}V}{dx^{2}}\right)
_{x=x_{\min }}/M}$=1.87~zs$^{-1}$. Curves for initial energies of $E_{tot}=0$
and $B/2$ are indicated for a temperature of $T$=1~MeV. The most noticeable
difference from the high friction results is the oscillations at small times
associated with the underdamped motion. Otherwise the general behavior is
similar. The calculation with $E_{tot}=0$, starting further from the barrier
energy, shows a suppression during the transient period while the
calculation with $E_{tot}=B/2$, starting closer to the barrier energy,
displays a fission enhancement.

The mean times $\tau _{\mathrm{MFPT}}$ and $\tau _{\mathrm{MLPT}}$ from
these simulations are displayed in Fig.~\ref{fig:underdamped}b. Because of
the underdamped motion, after the system crosses the saddle point ($x$=0) it
keeps going and never returns. Thus the mean FPT reaches its asymptotic
value at the saddle point. Unlike the mean times in Fig.~\ref{fig:mftp}a
calculated for the same temperature, but with high friction, the mean time
here shows a more significant dependence on the initial condition. However,
this will decrease for smaller temperatures. The greater dependence on the
initial condition is a direct consequence of the fact that the fraction of
fission events occurring during the transient time $t_{trans}$ is larger. In
the same vein, the transient times are relatively larger compared to $t_{f}$
for these low friction simulations. Fig.~\ref{fig:underdamped}c shows the
evolution of $t_{f}$ and $t_{trans}$ with temperature. Although
qualitatively similar to the high friction results in Fig.~\ref{fig:times},
the two times become quite similar in magnitude for $T\gtrsim B$, but still $%
t_{trans}<t_{f}$.

\section{DISCUSSION}

These calculations have emphasized that the nature of the fission transients
is very sensitive to the initial deformation of the compound nucleus. Given
this sensitivity, it is important to consider what determines these initial
conditions and whether they lead to a fission delay? Only a fission delay
will result in an increase in the predicted presaddle neutrons. If instead
one has a transient fission enhancement, then this will not help explain the
large numbers of prescission neutrons measured in experiments. As far as the
present calculations are concerned, time zero should correspond to the time
when shape fluctuations commence. As these are thermally driven, then the
initial deformation should correspond to the configuration at which a
significant fraction of the final excitation energy is dissipated. For
proton, $\alpha $-particle, or other light-ion induced reactions, one might
expect the dissipation of the initial kinetic energy to occur without any
significant change in the deformation of the target nucleus. The choice of
the ground-state or a spherical configuration as the starting value, thus
seems reasonable. On the other hand for fusion reactions between heavy
fragments, a more deformed initial configuration may be appropriate. Clearly
the transient effects cannot be decoupled from the fusion dynamics.

In the \textsc{HICOL} dynamical model\cite{Feldmeier87}, which uses the full
wall friction\cite{Blocki78}, the fusion dynamics can be divided into two
time regimes. First an initial rapid dissipation of the relative kinetic
energy between the projectile and target and then a slower relaxation of the
shape degrees of freedom. For example, Ref.~\cite{Charity00} discusses a 
\textsc{HICOL} simulation for the $E/A$=5~MeV $^{64}$Ni+$^{100}$Mo ($J$=0)
reaction where the thermal excitation energy is dissipated in 0.5~zs leaving
a highly deformed mononuclear configuration whose subsequent evolution can
be treated in the limit of high friction. This deformed configuration would
represent the initial deformation in the Langevin simulations.

\begin{figure}[tbp]
\includegraphics*[scale=.4]{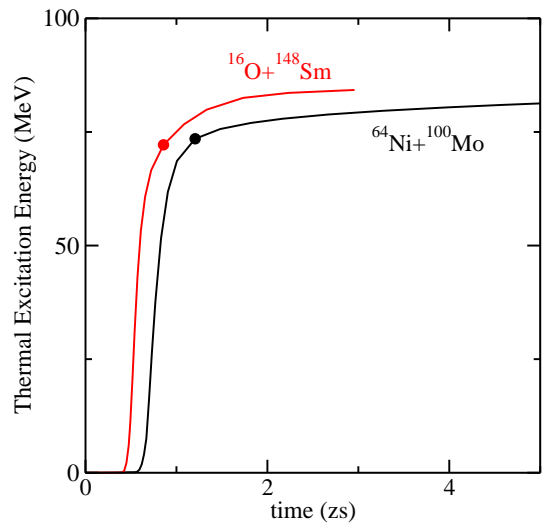}
\caption{(Color online) Predicted evolution with time of the dissipated
thermal excitation energy in the $E/A$=5 MeV $^{64}$Ni+$^{100}$Mo and the $%
E/A$=8.5 MeV$\ ^{16}$O+$^{148}$Sm reactions. Results were obtained with the 
\textsc{HICOL} code for $J$=40$\hbar $. The filled symbol indicates the
\textquotedblleft initial\textquotedblright\ configurations for which the
shapes are shown in Fig.~\protect\ref{fig:shape_40}.}
\label{fig:heat_40}
\end{figure}

\begin{figure}[tbp]
\includegraphics*[scale=.4]{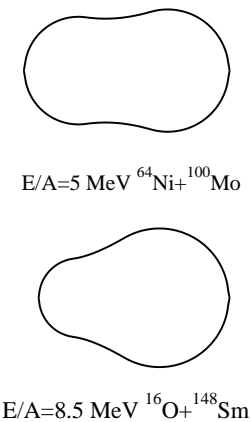}
\caption{Predicted \textquotedblleft initial\textquotedblright\ deformations
of the $^{164}$Yb compound nuclei ($J$=40~$\hbar $) predicted by the HICOL
code for the two indicated reactions.}
\label{fig:shape_40}
\end{figure}

Another example of this is illustrated in Fig.~\ref{fig:heat_40}. Here the
predicted dissipated thermal excitation energy is shown as a function of
time for two reactions ($^{64}$Ni+$^{100}$Mo and $^{16}$O+$^{148}$Sm) making
the same $^{164}$Yb compound nucleus with excitation energy of 100 MeV and $%
J $=40~$\hbar $. Both reactions show a rapid rise in the thermal energy over
a period less than 1~zs and subsequently this energy plateaus. The
difference between the plateau values in the two reactions reflects the
difference in the deformation-plus-rotational energy of the initial
configurations. The times indicated by the filled symbols in Fig.~\ref%
{fig:heat_40} are at the end of the periods of rapid dissipation and will be
taken as the \textquotedblleft initial\textquotedblright\ configurations
after which shape fluctuations need to be considered. The shapes of the
compound nucleus at these times are shown in Fig.~\ref{fig:shape_40} for the
two reactions. It is clear that the initial shape depends of the
entrance-channel mass asymmetry. However, it also depends of the bombarding
energy and impact parameter. The more symmetric the entrance channel, the
larger is the initial deformation. Thus it is possible that the nature of
the fission transients may change from a fission suppression for very
asymmetric entrance channels to an enhancement for the symmetric entrance
channels.


The assumption of high friction is not universal, e.g. Nix and Sierk have
used a reduced wall friction (scaled by 27\%) to explain the widths of the
giant isoscaler quadrupole and octupole resonances\cite{Nix86}. This reduced
friction is also consistent with experimental fission-fragment kinetic
energies. In this case, the damping of the fission coordinate is closer to
critical damping. With this friction, if the fusion dynamics bring in
sufficient energy into the motion of the fission coordinate (either
potential or kinetic), transient fission enhancement will again be possible.

Whatever the nature of the friction, a full understanding of the transients
would require consideration of more than just a fission coordinate. Initial
deformations for asymmetric entrance channels may have a mass asymmetry or
an octupole moment as is the case for the $^{16}$O+$^{148}$Sm reaction in
Fig.~\ref{fig:shape_40}. Thus the transients may affect the mass
distribution of the fission fragments. Transient fission may be considered
intermediate between statistical fission and quasifission as the rapid time
scale of transient fission may not lead to an equilibrium of the angular and
mass-asymmetry coordinates. As such, transient fission may contribute to the
observed suppression of the residue yield, and the broadening of the
fission-fragment mass distribution, for more symmetric entrance channels\cite%
{Berriman01}. These are usually considered a consequence of
\textquotedblleft fusion inhibition\textquotedblright\ due to quasifission
competition, but transient fission enhancements for the more symmetric
entrances channels will lead to similar effects.

Apart from fusion reactions, transient effects may play a role in the
sequential fission of targetlike (TLF) and projectilelike fragments (PLF)
following deep inelastic interactions. After the deep inelastic interaction,
these fragments may process a strong deformation with the remnant of the
neck, which connected the TLF\ and PLF during their interaction, orientated
towards the other fragment. This remnant gives the fragments octupole
deformations and hence an emergent mass-asymmetry.

In $^{100}$Mo+$^{100}$Mo and $^{120}$Sn+$^{120}$Sn reactions at $E/A\sim $20
MeV, Stefanini \textit{et al}. found the sequential fission axis for
asymmetric fission of the PLF is preferentially aligned along the beam axis
such that the smaller fission fragment is emitted towards the TLF\cite%
{Stefanini95}. They suggest an explanation based on transient fission to
explain their observation; due to the neck remnant, the PLF is created with
a deformation close to the saddle point for the mass asymmetric fission.
Thus there will be a transient enhancement for these mass asymmetries
allowing them to compete more favorably against evaporation and symmetric
fission. More recently, Hudan \textit{et al. }have used Langevin simulations
to illustrate these general ideas for more asymmetric splits\cite{Hudan04}.

\section{CONCLUSION}

One-dimensional Langevin simulations were performed to emphasize the strong
sensitivity of fission transients to the assumed initial shape distribution
of the compound nuclei. Fission delays or transient fission suppressions are
found if the assumed initial deformation of the compound nucleus is
spherical or near the ground-state value. However with an initial
fissionlike deformation for strong friction or with sufficient energy in the
fission degree of freedom for weak friction, then a transient fission
enhancement will result. It is argued that the initial conditions are
determined by the fusion dynamics and thus fission transients are dependent
on the entrance channel. The nature of the transients may change from a
suppression to an enhancement as the entrance-channel changes from
asymmetric to symmetric. Calculations which invoke fission delays to explain
the large number of prescission neutrons measured in experiments should be
reexamined in light of these considerations.

For temperatures smaller than the value of the fission barrier, the
probability of fissioning during the transient period is small. Thus if
there are no competing decay mode and all systems eventually fission, then
the mean first and last passage times for deformations beyond the saddle
point will be insensitive to the initial conditions as most fissions occurs
when the quasistationary rate is attained. Transient fission will only be
important when there is strong competition from evaporation where the only
systems that fission at all are those that fission early. Calculations
including strong competition from evaporative decays show a pronounce
dependence of the mean first and last passage times on the initial condition.

\begin{acknowledgments}
I wish to acknowledge many informative discussions with Professor L. Sobotka
and Dr. A. Sierk. This work was supported by the Director, Office of High
Energy and Nuclear Physics, Nuclear Physics Division of the U.S. Department
of Energy under contract number DE-FG02-87ER-40316.
\end{acknowledgments}

\end{document}